\documentclass[12pt]{article}
\usepackage{amssymb}
\usepackage{graphicx}
\begin{document}
\begin{titlepage}
\hskip 11cm \vbox{
\hbox{Bicocca--FT--99--40}
\hbox{BARI--TH 370/99}
\hbox{IFUP--TH 67--99}
\hbox{UNICAL--TH 99/6}}
\vskip 1cm
\centerline{\bf TOPOLOGY IN 2D CP$^{N-1}$ MODELS ON THE LATTICE:}
\centerline{\bf A CRITICAL COMPARISON OF DIFFERENT COOLING TECHNIQUES$^{~\ast}$}
\vskip 0.8cm
\centerline{B. All\'es$^{a~\dagger}$, L.~Cosmai$^{b~\ddagger}$,
M.~D'Elia$^{c~\P}$ and A.~Papa$^{d~\S}$}
\vskip .3cm
\centerline{\sl $^{a}$ Dipartimento di Fisica, Universit\`a di Milano--Bicocca and}
\centerline{\sl INFN--Sezione di Milano, I--20133 Milano, Italy}
\vskip .1cm
\centerline{\sl $^{b}$ INFN--Sezione di Bari, I--70126 Bari, Italy}
\vskip .1cm
\centerline{\sl $^{c}$ Dipartimento di Fisica, Universit\`a
di Pisa and INFN--Sezione di Pisa, I--56127 Pisa, Italy}
\vskip .1cm
\centerline{\sl $^{d}$ Dipartimento di Fisica, Universit\`a della Calabria 
and INFN--Gruppo collegato di Cosenza,}
\centerline{\sl I--87036 Arcavacata di Rende (Cosenza), Italy}
\vskip 1cm
\begin{abstract}
   Two-dimensional CP$^{N-1}$ models are used to compare
   the behavior of different cooling techniques on the lattice.
   Cooling is one of the most frequently used tools to study on the lattice
   the topological properties of the vacuum of a field theory.
   We show that different cooling methods behave in an equivalent
   way. To see this we apply the cooling methods 
   on classical instantonic configurations and on configurations 
   of the thermal equilibrium ensemble. We also calculate the 
   topological susceptibility by using the cooling technique.
\end{abstract}
\vfill

\hrule
\vskip.3cm
\noindent
$^{\ast}${\it Partially supported by MURST and by 
EC TMR program ERBFMRX-CT97-0122.}
\vfill
$ \begin{array}{ll}
^{\dagger}\mbox{{\it e-mail address:}} &
 \mbox{\tt alles@sunmite.mi.infn.it}\\
^{\ddagger}\mbox{{\it e-mail address:}} &
  \mbox{\tt cosmai@ba.infn.it}\\
^{\P}\mbox{{\it e-mail address:}} &
  \mbox{\tt delia@mailbox.difi.unipi.it}\\
^{\S}\mbox{{\it e-mail address:}} &
  \mbox{\tt papa@cs.infn.it}
\end{array}
$
\vfill
\vskip .1cm
\vfill
\end{titlepage}
\eject

\section{Introduction}
\label{sec:intro}

Two--dimensional (2d) CP$^{N-1}$ models~\cite{DDL78,Wit79} 
play an important role in quantum field 
theory since they represent a convenient theoretical laboratory to investigate 
analytical and numerical methods to be eventually exported to QCD. This is 
believed to be possible since these models
have several important properties in common with QCD, for instance 
asymptotic freedom, spontaneous mass generation, confinement and topological 
structure of the vacuum~\cite{Act85}. In particular, the CP$^{N-1}$ vacuum, 
like the QCD one, 
admits instanton classical solutions. The non--trivial structure of the vacuum
for the CP$^{N-1}$ models, as well as for QCD, can be probed by defining suitable 
operators, like the topological charge density and its correlators, and by
applying the prescriptions of quantum field theory. 
   Yet most of the observables relevant for the investigation of 
   the vacuum topological 
structure of a theory are beyond the reach of perturbation 
   theory. Therefore some approximation is needed which works also in the 
   non--perturbative regime. So far, the most effective and comprehensive 
   tool is represented by Monte Carlo simulations on a space--time lattice.
   The lattice approach for the study of the vacuum structure brings about 
   a very delicate task, that of separating
the quantum fluctuations at the level of the lattice spacing from the relevant
long--distance topological ones in any given configuration of the thermal 
equilibrium ensemble. Indeed, trying to assign a topological charge to a thermal 
equilibrium configuration, treating on the 
same footing short--distance (quantum) 
fluctuations and long--distance (topological) 
ones, would even prevent the reaching of the continuum limit. One powerful
technique, frequently adopted in the literature to study the topological 
structure of the vacuum, is the so--called cooling method~\cite{Tep86}.
The idea behind the cooling method is that the quantum fluctuations at the
level of the lattice spacing of a given configuration can be smoothed out
by a sequence of local minimizations of the lattice action, thus leading to a 
configuration where only the long--distance, topologically relevant fluctuations
survive. When this procedure is performed on a set of well--decorrelated 
equilibrium configurations, it yields 
a ``cooled'' equilibrium ensemble on which the expectation values of 
topological quantities can be determined. The precise definition of a cooling 
procedure is however arbitrary to a large extent: the lattice action to be
locally minimized can be chosen differently from 
the one used in the thermalization,
the number of cooling steps is a free variable, driving or control
parameters of the cooling can be introduced, and so on. Comparing the 
behavior of different cooling methods is an interesting issue,
especially when a new cooling strategy is proposed. 

The aim of the present work is to get insight into a new cooling method 
first adopted in Ref.~\cite{PPS99} for the SU(2) gauge theory and 
to compare it with two other methods, already known 
in the literature, namely the ``standard''~\cite{Tep86} cooling and 
its controlled version adopted by the Pisa group (see for instance 
Ref.~\cite{CDPV89}). The new cooling method, as it will be pointed out in 
the next Section, follows a different strategy with respect to the 
other two and could possess various interesting features, 
inaccessible to the other two cooling methods, like for instance preserving 
instanton--anti-instanton (I--A) 
pairs with interaction energy smaller than a fixable 
value~\cite{PPS99}.
The test--field for this comparison will be provided by the CP$^{N-1}$ models,
which have all the interesting topological properties of the 
SU($N$) gauge theories,
but are much easier to simulate on a lattice.

The paper is organized as follows: in Section~\ref{sec:def} we introduce 
the lattice action and the lattice discretization of the relevant topological 
quantities. In Section~\ref{sec:cooling} we describe in detail the three 
cooling methods. In Sections~\ref{sec:1-inst} and \ref{sec:ia}
we discuss their performance on 1--instanton and I--A classical
lattice configurations 
respectively.
In Section~\ref{sec:thermal} we consider their behavior on thermal equilibrium 
configurations and find a correspondence between the three cooling techniques.
In Section~\ref{sec:topsusc} we determine the physical value 
of the topological susceptibility of CP$^9$ using the new cooling method and 
compare the results to those from an alternative approach not based on cooling. 
Finally in Section~\ref{sec:summary} we draw our conclusions.  

\section{Lattice definition of action and observables}
\label{sec:def}

The CP$^{N-1}$ models describe the physics of a gauge invariant 
theory of interacting classical spins.
The continuum action for the 2d CP$^{N-1}$ model is 
\begin{equation}
S = \frac{1}{g} \int d^2x \, \overline{D_\mu z} \cdot D_\mu z\;,
\;\;\;\;\; D_\mu=\partial_\mu + i A_\mu\;,
\end{equation}
where $g$ is the coupling constant and $z(x)$ is a $N-$component complex
scalar field which obeys 
the constraint $\bar z(x) \cdot z(x) = 1$.  The bar over
a complex quantity means complex conjugate and a central dot among
vectors implies the scalar product: 
$\bar z(x) \cdot z(x)\equiv \sum_{i=1}^N \bar z_i(x) z_i(x)$. 

We regularize the theory on a square lattice using the standard discretization:
\begin{equation}
S^L = -N\beta\sum_x\sum_{\mu=1,2}
   \biggl(\bar z(x+\widehat\mu)\cdot z(x) \,\lambda_\mu(x) +
   \bar z(x) \cdot z(x+\widehat\mu)\, \bar\lambda_\mu(x) - 2\biggr)\;,
\end{equation}
where $x$ indicates a generic site 
of the lattice; $\lambda_\mu(x)$ is a U(1) gauge field 
($\bar\lambda_\mu(x)\,\lambda_\mu(x) =  1$) and $\beta\equiv1/(N g)$
where $g$ is the bare lattice coupling constant.
We used the standard action both to generate thermal equilibrium configurations
in Monte Carlo simulations and during the cooling procedure. 

The continuum topological charge is defined as
\begin{equation}
Q \equiv \int d^2x \, q(x) \equiv 
\frac{i}{2\pi} \int d^2x \, \epsilon_{\mu\nu} \overline{D_\mu z} 
\cdot D_\nu z = \frac{1}{2\pi}\oint dx_\mu A_\mu \;,
\label{q:continuum}
\end{equation}
where $\epsilon_{\mu\nu}$ is the antisymmetric tensor with
$\epsilon_{12}=1$ and
$q(x)$ is the topological charge density which can be written as the 
divergence of a topological current 
$K_\mu(x)\equiv1/(2\pi)\epsilon_{\mu\nu}A_\nu$,
$q(x) =\partial_\mu K_\mu(x)$~\cite{DDL78,Hoo76}.
The last integral in Eq.~(\ref{q:continuum}) is calculated on the 2d
plane along the circle of infinite radius.

The topological susceptibility $\chi$ is a renormalization group
invariant quantity 
which gives a measure of the amount of topological excitations in the
vacuum. It is defined as the two--point zero--momentum correlation of
the topological charge density operator $q(x)$, 
\begin{equation}
\label{chi}
\chi \equiv \int d^2x \, \langle 0 | \, T\, [q(x) q(0)] \, | 0 \rangle\;.
\end{equation}
This equation needs a prescription for the contact term coming from
the product of operators at the same point. We use the
prescription~\cite{Cre79}
\begin{equation}
\label{prescr}
\chi = \int d^2x \,  \partial_\mu
\langle 0 | \, T\, [K_\mu(x) q(0)] \, | 0 \rangle\; , 
\end{equation}
which corresponds to the physical requirement that $\chi = 0$ in the sector
with trivial topology (see for instance~Ref.~\cite{Meggiolaro:1998}).

On the lattice any discretization of the 
topological charge density with the correct na\"{\i}ve continuum limit
can be used.
A standard definition is
\begin{equation}
q^L(x) \!\equiv \!-{i\over 2\pi}\sum_{\mu\nu} \epsilon_{\mu\nu}
   {\rm Tr}\left[ P(x)\Delta_\mu P(x)
   \Delta_\nu P(x) \right]\;,
\label{q^L}
\end{equation}
with
\begin{equation}
\Delta_\mu P(x) \equiv \frac{P(x{+}\widehat\mu) - 
 P(x{-}\widehat\mu)}{2}~, \;\;\;\;\;
P_{ij}(x) \equiv \bar z_i(x) z_j(x) \;.
\label{defP}
\end{equation}

The lattice topological susceptibility is defined as
\begin{equation}
\label{chi^L}
\chi^L \equiv \biggl\langle \sum_x q^L(x)q^L(0) \biggr \rangle =
{1\over L^2} \left< \left(Q^L\right)^2 \right>\;,
\end{equation}
where $Q^L\equiv\sum_xq^L(x)$ and $L$ is the lattice size in lattice units.
Unless otherwise stated, hereafter the brackets $\langle\cdot\rangle$
mean average over decorrelated equilibrium configurations.

In the CP$^{N-1}$ model $q(x)$ is a renormalization group invariant
operator. Imposing that $q(x)$ does not renormalize in
the continuum generally implies a finite multiplicative
renormalization $Z$ on the lattice~\cite{CDP88}
\begin{equation}
\label{z}
q^L(x) = a^2 \, Z(\beta)\, q(x) + O(a^4)\;,
\end{equation}
$a$ being the lattice spacing.

Moreover, the lattice topological susceptibility $\chi^L$
in general does not meet the continuum prescription given 
in Eq.~(\ref{prescr}),
thus leading to an additive renormalization which consists in 
mixings of $\chi^L$ with operators which have the same quantum numbers
as $\chi$,
\begin{equation}
\label{reneq}
\chi^L(\beta) = a^2 Z(\beta)^2\chi + M(\beta)\;,
\end{equation} 
where $M(\beta)$ indicates such mixings.
The extraction of $\chi$ from the lattice thus requires the determination
of both the multiplicative and additive renormalizations, $Z(\beta)$ and
$M(\beta)$. This is a feasible task, as will be shown in 
Section~\ref{sec:topsusc}, and 
in this way one obtains the so--called ``field theoretical'' determination
of the topological susceptibility.

Alternatively one can use cooling to determine the topological
susceptibility. This method will be described extensively
in the next Section. The idea is that, by an iterative process of
local minimization of the action, the quantum fluctuations at the scale
of the ultraviolet cut--off, which are responsible for the renormalizations,
are suppressed thus implying 
$Z \to 1$ and $M \to 0$ in Eq.~(\ref{reneq}) as the cooling iteration
goes on, so that
$a^2\,\chi$ can be extracted directly from the measurement of
$\chi^L$ after cooling, as it will be shown in Section~\ref{sec:topsusc}.

\section{The cooling method}
\label{sec:cooling}

A cooling step applied on a lattice configuration consists 
in assigning to every field variable a new value chosen in order to
locally minimize the lattice action.
The iteration of this procedure converges to a configuration which should 
represent a solution of the Euclidean classical equations of motion.
In the continuum 2d CP$^{N-1}$ models, as well as in the case of 
4d SU($N$) gauge theories, there are 
classical solutions that belong to a definite topological charge sector, 
identified by an integer value of the topological charge $Q$ and 
by a finite action ($S = 2 \pi |Q|/g$ in the case of CP$^{N-1}$). 
The classical solutions with $Q\neq0$ 
are called instantons or anti-instantons (depending on the sign of
$Q$). On the lattice, owing to the 
artifacts introduced by the discretization of the theory,
configurations with non--trivial topology can represent 
only approximated solutions
of the lattice equations of motion. The common belief is that a moderately long 
cooling sequence can be able to wash out the quantum fluctuations of a given
lattice configuration, so revealing the underlying metastable state
with possible non--trivial 
topology, and that this state is the lattice counterpart 
of a continuum classical one. Of course, if the cooling is further 
protracted, the metastable state will eventually ``decay'' into the trivial vacuum.
Stated in other words, once the quantum fluctuations have been eliminated and
the topological bumps of a configuration (instantons or anti-instantons) have
been revealed, if the cooling is further 
protracted then the isolated instantons or 
anti-instantons shrink, the I--A pairs annihilate and the
configuration falls into the trivial vacuum of the lattice theory.
In consideration of the above arguments, it is clear that a great care is needed 
in determining when a cooling iteration has to be stopped. Nevertheless,
there is no way to prevent the loss of topological
signal at lattice scales of the order of the lattice spacing, 
since topology at these 
scales cannot be distinguished from quantum fluctuations. It must be pointed out
however, that any loss of topological signal due to cooling which occurs at 
fixed scales in {\em lattice units} becomes irrelevant in {\em physical units} 
as far as the coupling is tuned towards its critical value, i.e. in the continuum
limit, when the lattice spacing $a$ tends to zero. 
This is not the case only when the instanton size distribution is 
singular in the short--distance regime, as in the 2d 
CP$^1$ model~\cite{FP94,BBHN96}.

In the case of the 2d CP$^{N-1}$ models, the cooling algorithm amounts to
replacing at each lattice site $x$ 
the field variables $z(x)$ and $\lambda_\mu(x)$
by new ones $z^{\mathrm{new}}(x)$ and $\lambda_\mu^{\mathrm{new}}(x)$,
chosen in order to locally
minimize the lattice action, while the other field variables are left
unchanged. A cooling step consists in a sweep through the entire
lattice volume in order to apply this 
local minimization sequentially at every  site
$x$. The terms in the action density which depend on
$z(x)$ and $\lambda_\mu(x)$ for a given lattice site $x$ have the form
\begin{equation}
\label{sz}
s_z = - 2 \beta N \: {\rm Re} \{ \bar z(x) \cdot F_z(x)\}\;, \;\;\;\;
F_z(x) \equiv  \sum_{\mu=1,2} \biggl( z(x-\widehat\mu) \, 
 \lambda_\mu(x-\widehat\mu)
+ z(x+\widehat\mu) \, \bar \lambda_\mu(x) \biggr)\;,
\end{equation}
and
\begin{equation}
\label{slambda}
s_\lambda = - 2 \beta N \: {\rm Re} \{ \bar \lambda_\mu(x) \, F_\lambda(x,\mu) \}
\;,  \;\;\;\;\;\;\;\;\;\;
F_\lambda(x,\mu) \equiv  \, \bar z(x) \cdot z(x+\widehat\mu)\;,
\end{equation}
where $F_z(x)$ and $F_\lambda(x,\mu)$ are the respective forces.
It is easy to see that the local minimum of the lattice action is obtained
when the new field variables coincide with the corresponding
normalized forces:
\begin{equation}
\label{minimum}
z^{\mathrm{new}}(x) = { F_z(x)\over || F_z(x)||} \;, \;\;\;\;\;\;\;\;\;\;
\lambda_\mu^{\mathrm{new}}(x) =  {F_\lambda(x,\mu)\over 
 ||F_\lambda(x,\mu)||}\;, \;\;\;\;\;\;\;\;\;\;
 \mu=1,2 \, ,
\end{equation}
with $|| u ||\equiv \sqrt{\bar{u}\cdot u}$ for a generic complex
number or $N$--component vector $u$ ($z(x)$, $F_z(x)$,
$\lambda_\mu(x)$, $F_\lambda(x,\mu)$, etc.). 
According to the supplementary conditions under which the above replacements
are performed, we have different cooling procedures. 
The first cooling procedure, which we call ``standard'', is the one where 
the replacements $z(x) \rightarrow z^{\mathrm{new}}(x)$ and 
$\lambda_\mu(x) \rightarrow \lambda_\mu^{\mathrm{new}}(x)$
are unconstrained. The second one, which we call ``Pisa cooling'',
is the one where the replacement of $z(x)$ and $\lambda_\mu(x)$ is
performed if the distance between the old and the new fields is {\em smaller}
than a previously fixed parameter $\delta_{\rm Pisa}$: $||\lambda_\mu(x)-
\lambda_\mu^{\mathrm{new}}(x)|| \leq \delta_{\rm Pisa}$ and analogously
for the $z$ fields,
otherwise the new field is chosen to lie on the 
$(z(x),z^{\mathrm{new}}(x))$ ($(\lambda_\mu(x),\lambda_\mu^{\mathrm{new}}(x))$) 
plane, at a distance from $z(x)$ ($\lambda_\mu(x)$) exactly equal 
to $\delta_{\rm Pisa}$~\cite{CDPV89}. The 
third cooling procedure, called ``new cooling'' in the following, 
was introduced in Ref.~\cite{PPS99}. It consists in making the replacement only 
if the distance between the old and new fields
is {\em larger} than a fixed parameter $\delta$, 
otherwise the field is left unchanged.
We observe that the three methods follow a different strategy. In 
particular, while the Pisa cooling acts first on the smoother fluctuations
and the number of cooling iterations to be performed 
has to be fixed {\em a priori},
the new cooling performs local minimizations only if these fluctuations are 
larger than a given threshold and the iteration stops automatically when 
all the surviving fluctuations are smoother than the (lattice) threshold 
determined by the dimensionless parameter $\delta$. 
It is easy to see that the new field variables differ
from the old ones by terms which, in the continuum limit, are of order
$a^2$, where $a$ is the lattice spacing. 
Therefore the parameter $\delta$ 
is expected to scale in the continuum limit
like a quantity of dimension two.
\begin{figure}[tb]
\begin{minipage}{0.48\textwidth}
\begin{center}
\includegraphics[width=0.9\textwidth]{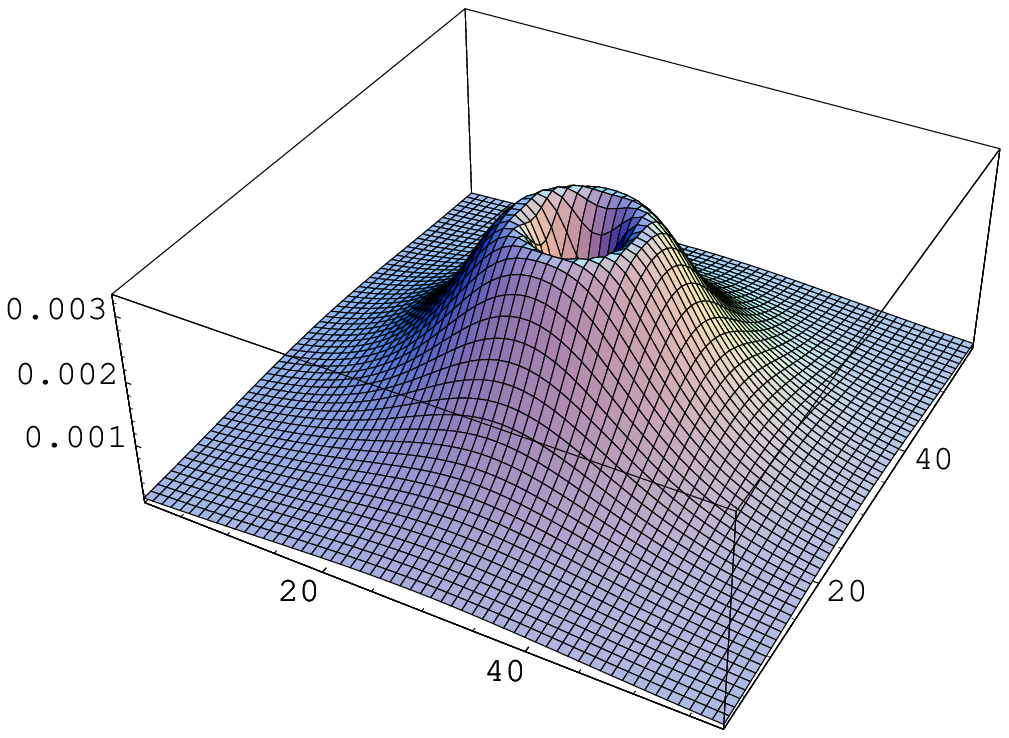}
\end{center}
\end{minipage}
\begin{minipage}{0.48\textwidth}
\begin{center}
\includegraphics[width=0.9\textwidth]{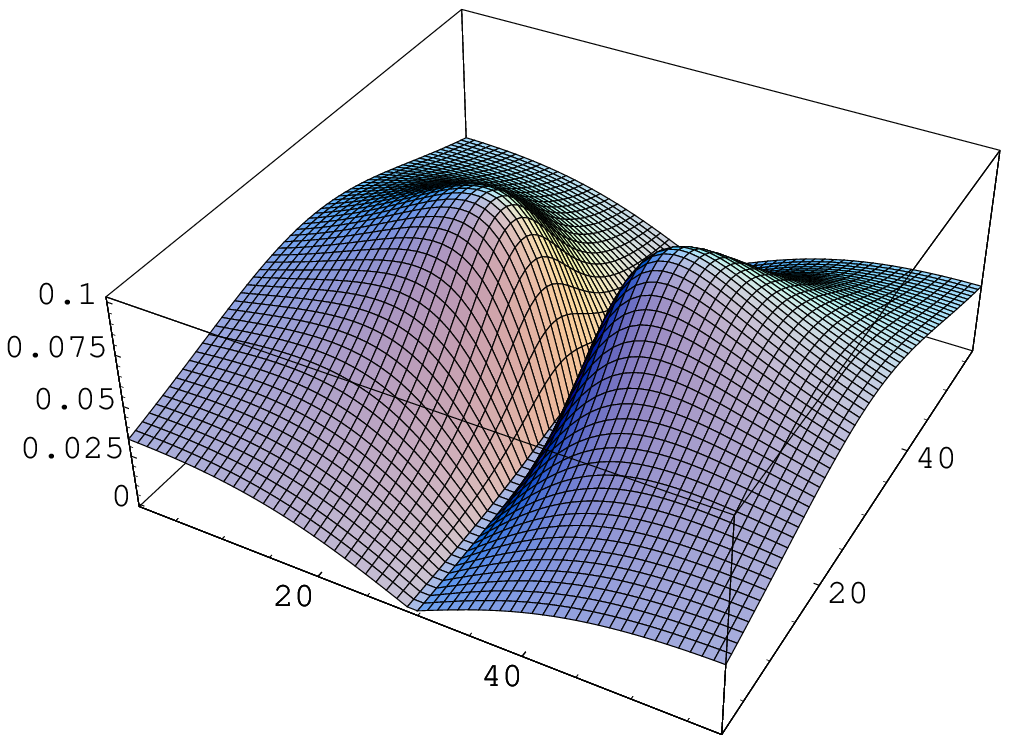}
\end{center}
\end{minipage}
\caption{$\theta_z(x)$ (left) and $\theta_\lambda(x,1)$ (right) for an
``artificial'' 1--instanton with size $\rho = 10 a$ placed in the middle of
a $60^2$ lattice.}\label{fig:1-inst}
\end{figure}
In the following we will compare these three cooling methods in several contexts,
namely on ``artificial'' 1--instanton configurations, on I--A pairs, 
on thermal equilibrium configurations and in the determination of the 
continuum topological susceptibility. Moreover we will establish a relation
between the number of cooling iterations in the standard and Pisa coolings 
with the parameter $\delta$ of the new cooling and will discuss the validity
of the picture of cooling as a diffusion process.

\section{Cooling 1--instanton configurations}
\label{sec:1-inst}

In the continuum CP$^{N-1}$ models, in the infinite 2d space--time,
a 1--instanton configuration is a classical solution of the equations of motion 
characterized by a topological charge one and a finite action ($S = 2\pi/g$). 
The explicit form of a continuum 1--instanton configuration is 
\begin{equation}
z(x) = \frac{w(x)}{||w(x)||}\;, \;\;\;\;\;
A_\mu = i \bar z(x) \cdot \partial_\mu z(x)\;,
\label{inst_cont}
\end{equation}
with
\begin{equation}
w(x) = u + \frac{(x_1-c_1) - i (x_2-c_2)}{\rho}\, v\;,
\end{equation}
where $u$ and $v$ are complex 
$N$--vectors which satisfy $ \bar u \cdot u = \bar v \cdot v 
= 1$ and $\bar u \cdot v = 0 $. The space--time coordinates $(c_1, c_2)$
are the center of the instanton and the real parameter $\rho$
represents a measure of its 
size. An ``artificial'' 1--instanton lattice configuration can be
represented 
by the discretization of the above continuum configuration on a 
2d finite volume lattice with periodic boundary conditions. In particular, we used 
the following expressions for the fields $z$ and $\lambda$~\cite{DDL78}
\begin{displaymath}
z_1(x) = \frac{\rho}{\sqrt{\rho^2 + (x_1-c_1)^2 + (x_2-c_2)^2}}\;,\;\;\;
z_2(x) = \frac{x_1-c_1 - i (x_2-c_2)}{\sqrt{\rho^2 + (x_1-c_1)^2 
+ (x_2-c_2)^2}}\;,
\end{displaymath}
\begin{eqnarray}
z_i(x) &=& 0\;, \;\;\;\;\; i=3,\ldots,N\;, \nonumber \\ 
\lambda_\mu(x) &=& 
\frac{\bar z(x+\widehat\mu) \cdot z(x)}{||\bar z(x+\widehat\mu) \cdot z(x)||}\;,
\end{eqnarray}
which correspond to choose $u=(1,0,0,\dots)$ and $v=(0,1,0,0,\dots)$.
Actually 1--instanton configurations cannot exist on a
2d torus even in the continuum~\cite{RR83} and they represent only approximate
solutions of the classical equations of motion. This approximation becomes better
and better as the ratio $\rho/(La)$ decreases.
On the lattice, this problem would lead to metastability even in the case
of an ideally perfect discretization. However the above--defined
artificial 1--instanton 
lattice configurations are enough for our purposes
since they will be used only as test configurations for cooling
methods and not to extract physical information.

\begin{figure}[tb]
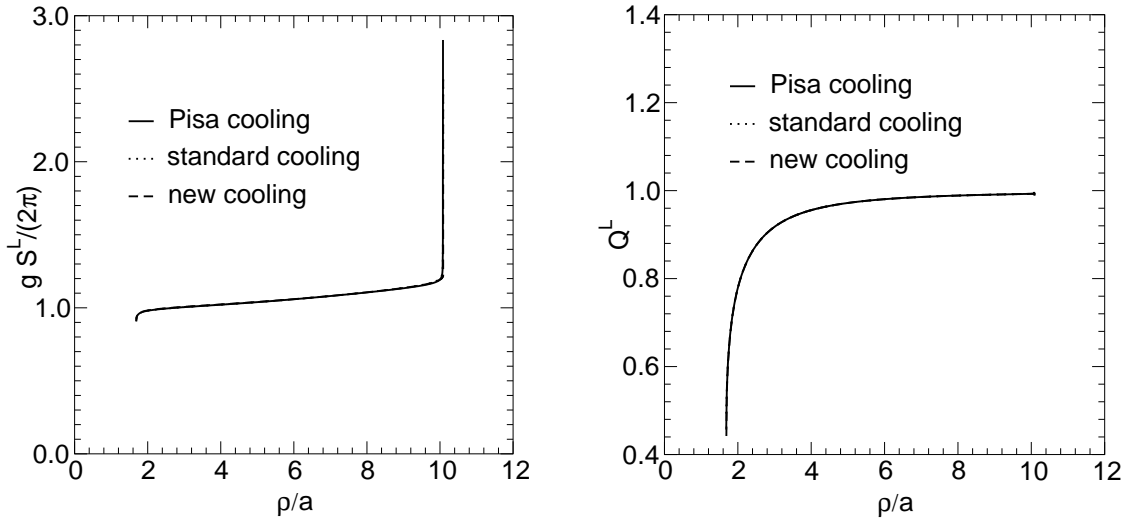

\begin{minipage}{0.48\textwidth}
\begin{center}
\includegraphics[width=0.9\textwidth]{action_rho.eps}
\end{center}
\end{minipage}
\begin{minipage}{0.48\textwidth}
\begin{center}
\includegraphics[width=0.9\textwidth]{q_rho.eps}
\end{center}
\end{minipage}
\caption{$g\,S^L/(2\pi)$ (left) and $Q^L$ (right) 
versus $\rho/a$ during a protracted
cooling iteration with the three cooling methods. The starting configuration
is a 1--instanton with size $\rho = 10 a$ placed in the middle of a $60^2$ 
lattice.}\label{fig:Q,S_vs_rho}
\end{figure}

In Fig.~\ref{fig:1-inst} we show the distribution of the angles $\theta_z(x)$ and 
$\theta_\lambda(x,1)$ for an artificial 1--instanton with size $\rho=10 a$
located in the middle of a $60^2$ lattice. These angles are defined as
\begin{equation}
||\lambda_\mu(x) - \lambda_\mu^{\rm new}(x)||\equiv 2 \sin
{\theta_\lambda(x,\mu) \over 2}\;,
\qquad\qquad\qquad
||z(x) - z^{\rm new}(x)||\equiv 2 \sin
{\theta_z(x) \over 2} \,.
\end{equation}

It is clear that, while the 
standard cooling acts indistinctly on each lattice site, the Pisa cooling and 
the new cooling deform 1--instanton lattice configurations starting from 
different regions of the lattice. The new cooling 
will act first on the region around the center of the instanton, 
the Pisa cooling on the border regions. Therefore it is not obvious
whether under the three coolings the configurations will 
look the same sweep after sweep. We did therefore the following 
check: we performed a long ($O(1000)$ steps) iteration of the standard and Pisa
coolings\footnote{For the Pisa cooling we always use $\delta_{\rm Pisa}=0.2$, 
following Ref.~\cite{CRV92}.} on an artificial 1--instanton
configuration with size $\rho=10a$ and after each sweep we extracted 
the topological 
charge $Q^L$, the action $S^L$ and the size $\rho/a$. This size was determined
by a very local procedure, namely by looking for the maximum of the lattice
action density $s^L_{\mathrm{max}}$ and using the relation 
$s^L_{\mathrm{max}}=2a^2/\rho^2$, valid for a 1--instanton
configuration. Then, starting
from the same configuration, we applied the new cooling for
$\delta$ as small as $5\times 10^{-4}$ and for each configuration obtained during
the cooling iteration, we determined again $Q^L$, $S^L$ 
and $\rho/a$. Finally, we put on a plot $Q^L$ and $S^L$ as functions
of $\rho/a$. The result is
shown in Fig.~\ref{fig:Q,S_vs_rho}: the values of $Q^L$ and $S^L$
during the three cooling methods lie on the same curve.
The same plot was obtained by using a different procedure to determine
the size of the cooled configurations, namely a global fit of the
lattice action density to the continuum expression for the action
density of a 1--instanton configuration, Eq.~(\ref{inst_cont}).
We interpret these results as an indication that the three coolings 
deformed in the same way the starting configuration. If this conjecture 
is correct, it should be possible to put into correspondence
the $\delta$ parameter of the
new cooling with the number of steps of the standard or Pisa coolings. 
This will  be done in Section~\ref{sec:thermal}.

\begin{figure}[tb]
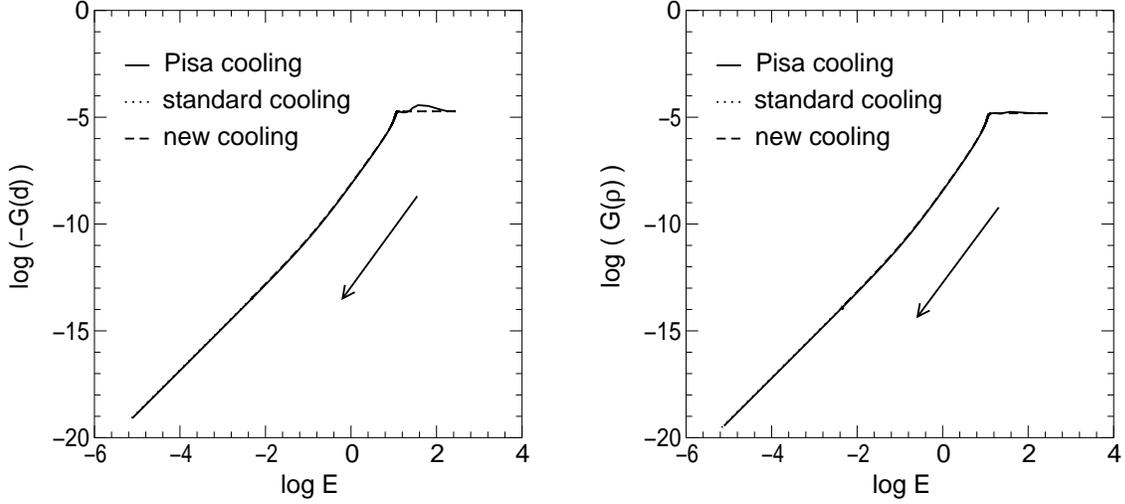

\begin{minipage}{0.48\textwidth}
\begin{center}
\includegraphics[width=0.9\textwidth]{ene-qcorr_dist_ia.eps}
\end{center}
\end{minipage}
\begin{minipage}{0.48\textwidth}
\begin{center}
\includegraphics[width=0.9\textwidth]{ene-qcorr_size_ia.eps}
\end{center}
\end{minipage}
\caption{${G}(d)$ (left) and
${G}(\rho)$ (right) versus the internal energy $E$
during a protracted cooling iteration with the three cooling methods
in a log--log scale. The 
starting configuration is an I--A pair with centers
$c^{(I)}=(15a,30a)$, $c^{(A)}=(45a,30a)$ (hence 
$d \equiv |c^{(I)} - c^{(A)}| = 30 a$) and $\rho = 5 a$ on a $60^2$ 
lattice. The arrows show the direction of the evolution under 
cooling.}\label{fig:Qcorr_vs_S}
\end{figure}

\section{Cooling I--A pairs}
\label{sec:ia}

An I--A configuration in the continuum 2d CP$^{N-1}$ models can be taken in 
the form given in Refs.~\cite{BL81,DM99},
\begin{equation}
z(x) = \frac{w(x)}{||w(x)||}\;, \;\;\;\;\;
A_\mu = i \bar z(x) \cdot \partial_\mu z(x)\;,
\end{equation}
with
\begin{eqnarray}
w_1(x_1,x_2) &=& (x_1+i x_2-a_1)(x_1-i x_2 - \bar b_1)\;, \nonumber \\
w_2(x_1,x_2) &=& (x_1+i x_2-a_2)(x_1-i x_2 - \bar b_2)\;, \nonumber \\
w_i &=&0 \;, \;\;\;\;\; i=3,\ldots,N\;.
\label{ia_cont}
\end{eqnarray}
In this expression, the complex numbers $a_{1,2}$ and $b_{1,2}$ are related
to the position of the center and the size of the instanton and
anti-instanton through the following relations:
\[
a_1 = c^{(I)}_1 + i c^{(I)}_2 - \rho^{(I)}\;, \;\;\;\;\;\;\;\;\;\;
a_2 = c^{(I)}_1 + i c^{(I)}_2 + \rho^{(I)}\;, \;\;\;\;\;\;\;\;\;\;
\]
\begin{equation}
b_1 = c^{(A)}_1 + i c^{(A)}_2 - \rho^{(A)}\;, \;\;\;\;\;\;\;\;\;\;
b_2 = c^{(A)}_1 + i c^{(A)}_2 + \rho^{(A)}\;, \;\;\;\;\;\;\;\;\;\;
\end{equation}
where $c_i^{(I,A)}$ is the $i$--th coordinate of the center of the
instanton or anti-instanton and $\rho^{(I,A)}$ are their sizes.
As ``artificial'' I--A lattice configuration we took
the discretization of the above continuum configuration on a 2d finite
volume 
lattice with periodic boundary conditions. For simplicity, we used $\rho^{(I)}=
\rho^{(A)}\equiv \rho$.
 
Starting from several I--A pairs with different values of the distance 
between the centers $c^{(I)}$ and $c^{(A)}$ and of the size $\rho$, we performed
long cooling sequences with the three methods, namely $O(1000)$ iterations
for the standard and Pisa coolings and $\delta$ as small as $5\times 10^{-4}$ for
the new cooling. In order to have an indication of how the cooling
process works, we examined by eye the distribution of 
the topological charge and action 
densities during the different cooling iterations.
The evolution of these densities looked the same, 
no matter which cooling procedure was adopted. Specifically, the
distribution of the action density shows at the beginning two equal
instanton bumps, which for $d\gg \rho$ ($d$ is the distance between
the two centers $c^{(I)}$ and $c^{(A)}$)
are well--separated; then, as the cooling goes on, these bumps lower in height
and merge together, up to complete annihilation. In order to make quantitative the 
statement that the three coolings behave essentially in the same way, we 
studied the shell correlation function of the topological 
charge density. This function is defined as 
${G}(r)\equiv 1/N_r\sum_{x,y} q^L(x) q^L(y)$
where the sum is extended over all pairs of sites $x$, $y$ which satisfy 
$r \leq |x-y| < r + \Delta\,r$ and $N_r$ is the number of those pairs for
a given value of $r$ (we have chosen $\Delta\,r=0.6 \, a$).
We determined ${G}(r)$ on every configuration resulting after 
each cooling iteration\footnote{\label{foot2}From reflection
positivity it follows that the two--point correlation function
of the topological charge density,
and consequently also $\langle G(r) \rangle$, 
is negative at distances $r > 0$. However reflection positivity is lost
during cooling, since cooling affects the quantum structure
of the theory leaving only the semiclassical background intact.
After cooling one in general obtains positive values for 
$\langle G(r) \rangle$, apart from those large separations
which the cooling procedure has not yet affected.}.
We compared the results which presented the same value for the
internal energy $E$ which is defined by
\begin{equation}
\label{energy}
E \equiv \frac{1}{2L^2} \, \sum_x\sum_{\mu=1,2}
  \left(\bar z(x+\widehat\mu)\cdot z(x) \, \lambda_\mu(x) +
   \bar z(x) \cdot z(x+\widehat\mu)\, \bar\lambda_\mu(x) - 2\right)\;,
\end{equation}
and is proportional to the lattice action.
In Fig.~\ref{fig:Qcorr_vs_S} we show the behavior of the shell correlation 
versus $E$ during the three (protracted) coolings for the case of $r$ 
equal to the distance $d$ between the centers of the instanton and of 
the anti-instanton (left) and to the instanton (or anti-instanton)
size $\rho$ (right). 
The three curves fall on top of each other, apart from a small deviation 
between the Pisa cooling and the other two at the very first (1--2) cooling 
steps, due to (unphysical) finite size effects. Since the shell correlation of the
topological charge density is related to the size of the instantons, the 
conclusion is that the three cooling methods perform equivalently
in deforming the size of both the instanton and the anti-instanton in
the I--A pair and in modifying the distance between
the two peaks.

\begin{figure}[tb]
\begin{center}
\hspace{0.3cm}
\includegraphics[width=0.65\textwidth]{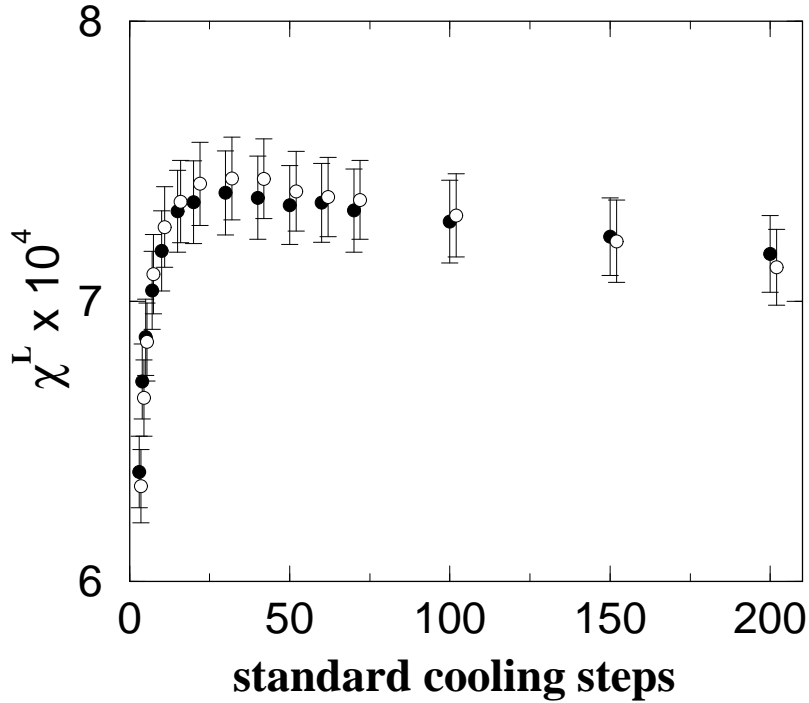}
\end{center}
\caption{Comparison of the determinations of $\chi^L$ after equivalent amounts
of new and standard coolings for the CP$^3$ model at $\beta=1.05$ on a
$76^2$ lattice. The new cooling results are reported versus
equivalent iterations of standard cooling steps following 
Table~1. Full (open) symbols indicate the new (standard) cooling.}
\label{fig:chiconfr}
\end{figure}

\begin{figure}[tb]
\begin{minipage}{0.52\textwidth}
\begin{center}
\includegraphics[width=0.9\textwidth]{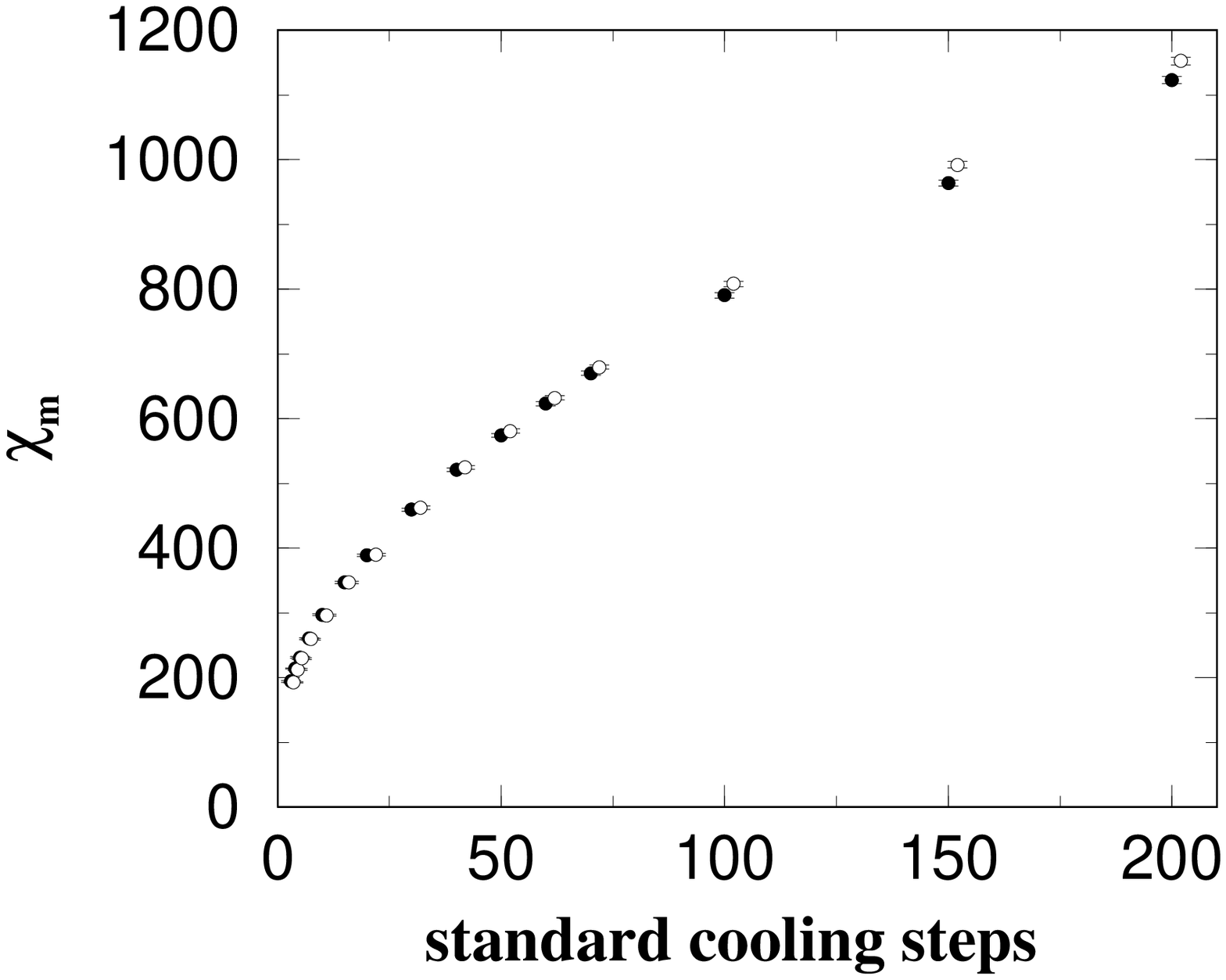}
\end{center}
\end{minipage}
\begin{minipage}{0.47\textwidth}
\begin{center}
\includegraphics[width=0.9\textwidth]{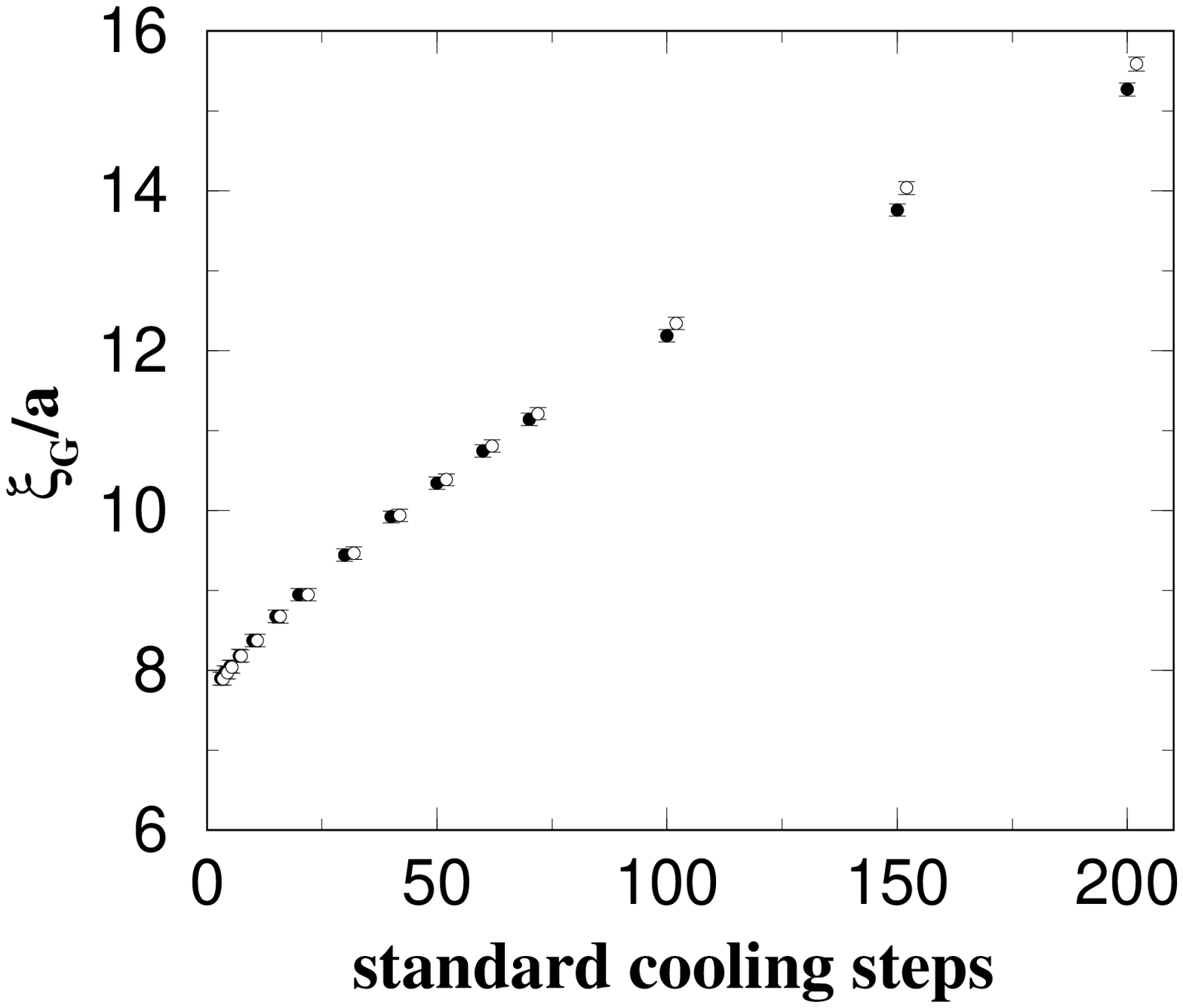}
\end{center}
\end{minipage}
\caption{The same comparison as in Fig.~\ref{fig:chiconfr} for the magnetic
susceptibility and the correlation length as defined in
Eq.~(\ref{csig}). Data points have been horizontally shifted for the
sake of readability. Full (open) symbols indicate the new (standard)
cooling.}
\label{fig:susc,csi}
\end{figure}

\begin{figure}[tb]
\begin{center}
\hspace{0.3cm}
\includegraphics[width=0.65\textwidth]{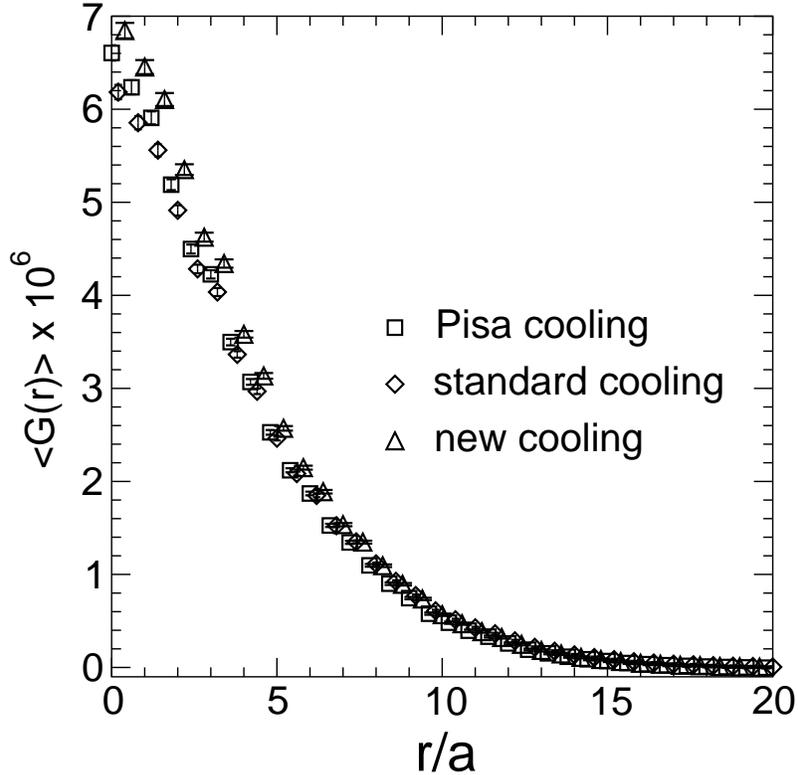}
\end{center}
\caption{Two--point shell correlation of the lattice topological charge density
operator ($r$ is the distance between two lattice sites)
after cooling for the CP$^3$ model at $\beta=1.05$ on a $76^2$
lattice. The data have been slightly horizontally shifted.}\label{fig:q_corr}
\end{figure}

\begin{figure}[tb]
\begin{minipage}{0.48\textwidth}
\begin{center}
\includegraphics[width=0.9\textwidth]{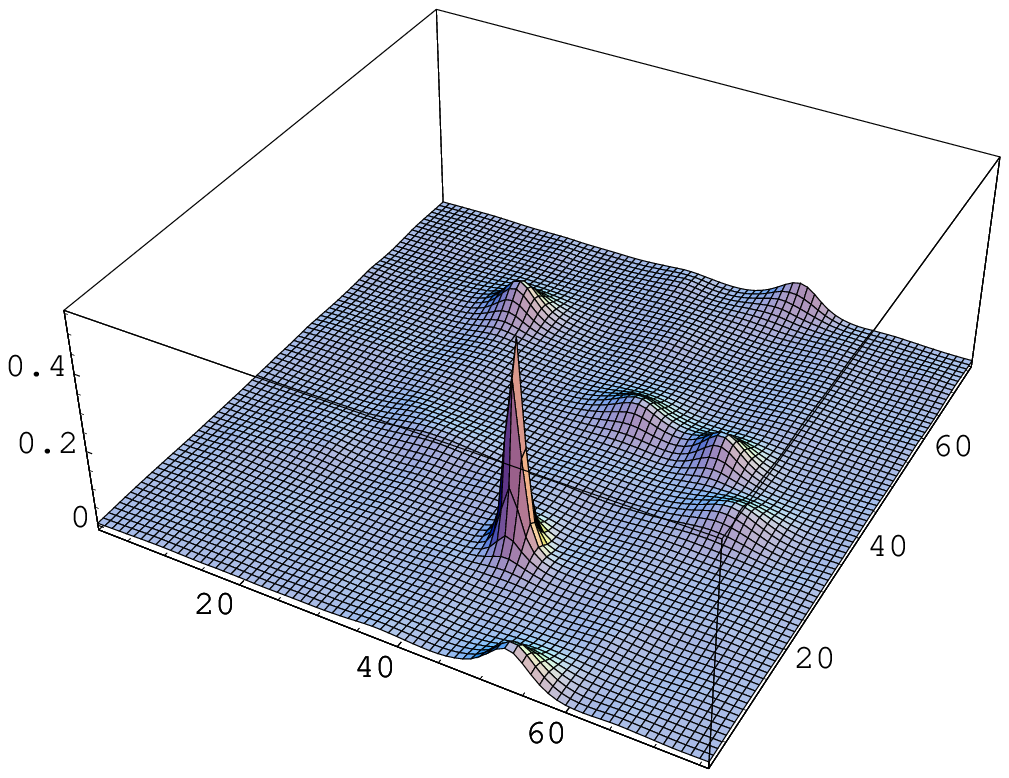}
\end{center}
\end{minipage}
\begin{minipage}{0.48\textwidth}
\begin{center}
\includegraphics[width=0.9\textwidth]{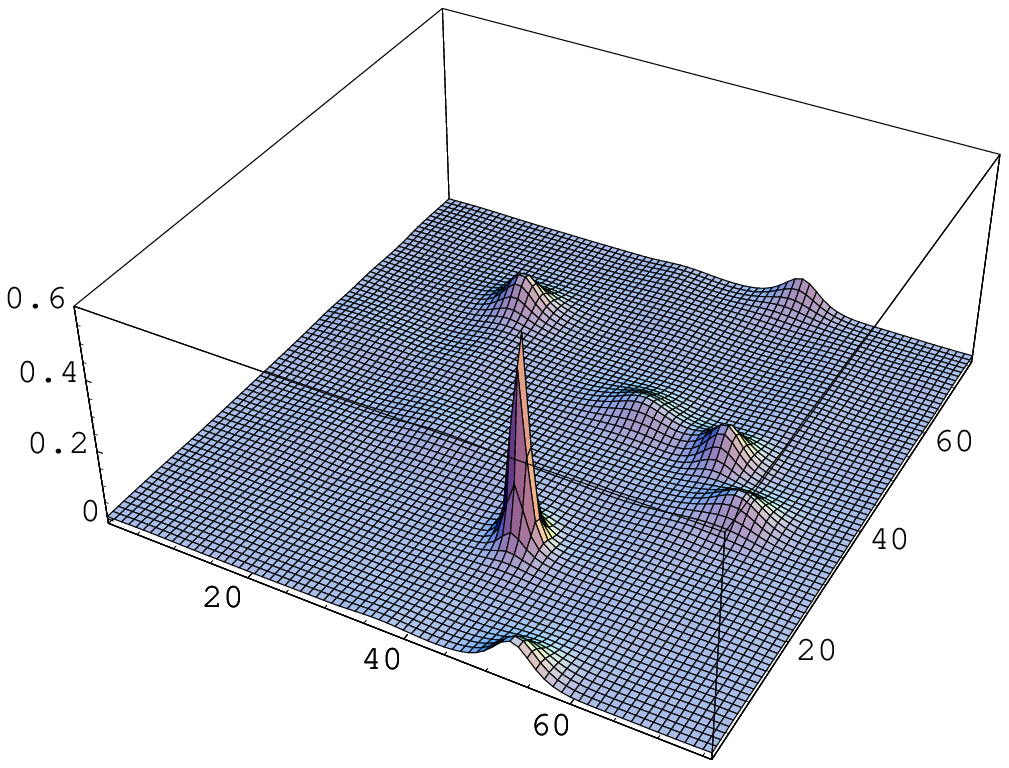}
\end{center}
\end{minipage}
\caption{Action density for a configuration obtained by cooling a thermal 
equilibrium configuration of CP$^3$ at $\beta=1.05$ on a $76^2$ lattice by 
the new cooling with $\delta=0.007$ (left) and by 30 iterations of the standard 
cooling (right). The two amounts of cooling are equivalent in the sense 
explained in the text.}\label{fig:therm}
\end{figure}

\begin{figure}[tb]
\begin{center}
\hspace{0.3cm}
\includegraphics[width=0.65\textwidth]{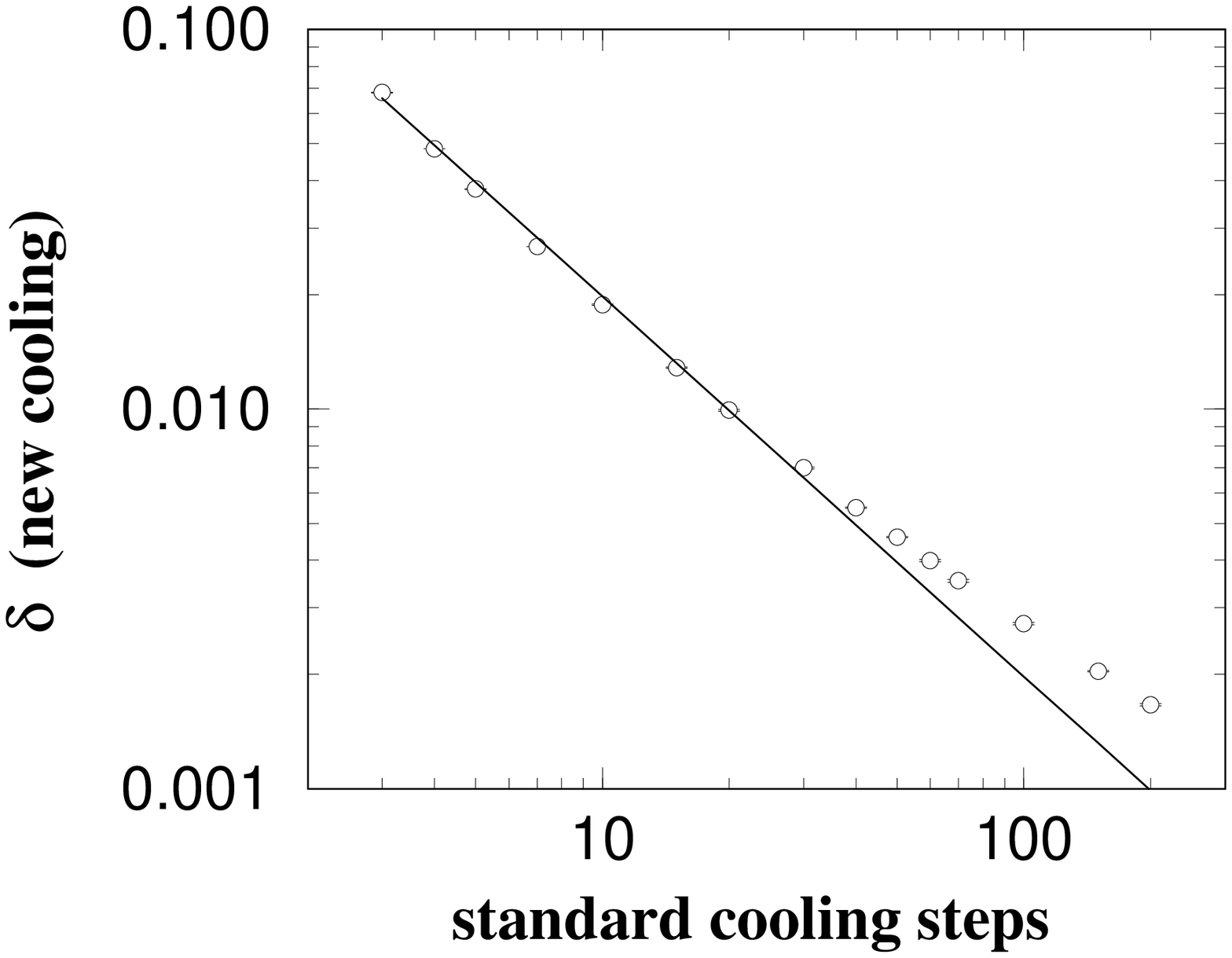}
\end{center}
\caption{Relation between the new cooling parameter $\delta$ and 
the number of standard cooling iterations $n$, at $\beta = 1.05$ 
on a lattice $76^2$ in the CP$^{3}$ model, together with a best fit
to a function proportional to $1/n$.}
\label{fig:deltan}
\end{figure}

\section{Cooling on the equilibrium ensemble}
\label{sec:thermal}

Using artificial configurations with non--trivial topology provides interesting 
suggestions about a cooling procedure, but in real life the cooling must be applied
on thermal equilibrium configurations. Among them there are 
configurations which contain several instanton and 
anti-instanton bumps, merged 
in a sea of quantum fluctuations. These configurations
can represent as well good test--fields for the different cooling procedures. 
Their evolution under cooling can be visualized by the distribution of
the action density or of the  
topological charge density. In principle we could expect 
that the three cooling 
methods described in Section~\ref{sec:cooling} 
make the starting thermalized configuration
evolve along different directions. In particular, for a not too protracted
cooling, one expects that all the three methods have been able to erase the quantum
fluctuations, but the number, type and location of the surviving topological
bumps in the cooled configuration can be rather different. In order to check
this expectation, it is necessary to find a criterion to make
a correspondence between the number of iterations in the standard or 
Pisa coolings with a $\delta$ parameter for the new cooling. 
We defined an effective temperature for configurations 
obtained during the cooling iteration in such a way that the
comparison could be made only between configurations at the same
temperature. The most natural ``thermometer'' is the internal energy
$E$, Eq.~(\ref{energy}),
since this is the quantity which is minimized during the cooling procedure.

We put into practice the above considerations in the case of the
CP$^3$ model. We generated by Monte Carlo a sample of equilibrium
configurations on a $76^2$ lattice at $\beta=1.05$.
As in Ref.~\cite{CRV92}, the simulation algorithm is, for every
updating step, a mixture of 4
microcanonical updates and 1 over heat--bath~\cite{petr}.
We measured the expectation value of $E$ on the cooled ensembles obtained by
the three cooling methods after several number of iterations (standard and Pisa
coolings) and for several values of $\delta$ (new cooling). By comparing the 
results and imposing that $E$ be the same, 
we obtained a correspondence table between the number of cooling 
iterations for the standard and Pisa coolings with 
a value of the $\delta$ parameter in
the new cooling. We found for instance that for the CP$^3$ model 
at the above values of $\beta$ and lattice size,
30 iterations of the standard cooling correspond (in the sense
of the average value of $E$) to $\delta=0.007$ for the new cooling 
and to approximately 33 iterations of the Pisa cooling.

In general the matching between the number of standard 
and Pisa cooling iterations is not as good as the 
one obtained between $\delta$ in the new cooling and 
a fixed number of iterations in the standard (or Pisa) cooling.
The reason is that in the second case one can tune a 
continuous parameter while in the first only integer jumps of the
parameter controlling the amount of cooling are allowed. For this 
reason we have mainly investigated the relation
existing between the standard and new coolings.
In Table~1 we show the values of $\delta$ (new cooling) found in
correspondence to several numbers of iterations of the standard cooling
for the CP$^3$ model at $\beta=1.05$ on a $76^2$ lattice. Notice
that a different correspondence table may be found for other values of
$N$ and $\beta$.

Using the correspondence table we can directly compare
the different coolings. As a first step, we have computed the average
values of some physical quantities on samples obtained after applying
equivalent amounts of cooling on a set of equilibrium configurations.
In Fig.~\ref{fig:chiconfr} 
we show the results for the lattice topological susceptibility
with the standard and new coolings. By using the relation given in Table~1,
we report both determinations in terms of the number of 
standard cooling iterations.
It clearly appears that, regarding to the determination of the topological
susceptibility, the two cooling techniques are completely equivalent.

Analogously in Fig.~\ref{fig:susc,csi} we compare the determinations
of the magnetic susceptibility $\chi_{\rm m}$ 
and of the correlation length $\xi_G$ defined by
\begin{eqnarray}
 \chi_{\rm m}&\equiv&{1\over L^2} \sum_{x,y}\langle \hbox{Tr} 
 P(x) P(y)\rangle_{\rm conn}\;, \nonumber \\
\left(\frac{\xi_G}{a}\right)^2 &\equiv& {1\over4\sin^2(\pi/L)} \,
\left[\frac{\widetilde G_P(0,0)}{\widetilde G_P(0,1)} - 1\right] \;,
\label{csig}
\end{eqnarray}
where
\begin{equation}
\label{fourier}
\widetilde G_P(k) \equiv \frac{1}{L^2} \sum_{x,y} \langle {\rm Tr} \: P(x)\, P(y) 
\rangle_{\rm conn}\exp \left[i \frac{ 2 \pi}{L} (x-y)\cdot k\right] \,,
\end{equation}
is the lattice Fourier transform of the correlation of two local 
gauge--invariant composite operators $P_{ij}(x)$ which have been defined in
Eq.~(\ref{defP}) (the subscript ``conn'' means connected Green
function). The results for $\chi_{\rm m}$ and $\xi_G/a$
from Fig.~\ref{fig:susc,csi} indicate that in this case the
new cooling and the standard cooling show a good agreement, apart from
a small discrepancy at large numbers of iterations. 

We have also studied the shell 
correlation of two lattice topological charge density operators, 
$\langle{G}(r)\rangle$, for $r/a$ going from 0 to 20, 
after equivalent amounts of the 
three coolings. In Fig.~\ref{fig:q_corr}
we compare the standard cooling (30 iterations) to
the new cooling ($\delta = 0.007$) and the Pisa cooling (33
iterations, $\delta_{\rm Pisa} = 0.2$)\footnote{See footnote \ref{foot2} 
in Section~\ref{sec:ia} about the positivity of $\langle{G}(r)\rangle$
after cooling.}.
The results are in good agreement, although a small
deviation can be observed at small distances.
The suggestion of this outcome is that the three coolings 
affect the instanton size distribution in the small distance region in a 
slightly different way; the effect is however modest.

Next we compared the three coolings on a subset of the
above--generated thermal equilibrium ensemble for the CP$^3$ model.
To this aim, we have chosen from the thermal 
ensemble several configurations showing non--zero topological charge after 
30 iterations of the standard cooling and compared their action density and
topological charge density distributions by eye 
to those of the same configurations cooled by 
33 iterations of the Pisa cooling and by the new cooling with $\delta=0.007$. 
For all the configurations considered, we observed that the bumps corresponding 
to instantons and anti-instantons had the same shape, number and location 
(see Fig.~\ref{fig:therm} for an example). 

These considerations provide evidence that the three
cooling methods defined in Section~\ref{sec:cooling} behave essentially in an 
equivalent way, not only on average, but also configuration by configuration, 
and that they can be related by a simple correspondence between
number of iterations (standard and Pisa coolings) and $\delta$ (new cooling).

We close this Section with some considerations about 
the usual picture of cooling as a diffusion process. 
As we pointed out in Section~{\ref{sec:cooling}}, 
the parameter $\delta$ in the new cooling behaves, in the continuum limit, 
like a physical quantity of dimension two. Therefore, indicating with 
$r_0$ the physical scale up to which 
cooling affects the quantum fluctuations and following Ref.~\cite{PPS99},
we infer that $r_0 \propto \delta^{-1/2}$ in the continuum limit.
On the other hand, the standard cooling is usually believed to act
as a diffusion process, so that, if $n$ is the number of iterations,
it should affect fluctuations up to a scale $r_0 \propto n^{1/2}$.
Using the relation found between $n$ and $\delta$, we can check 
both these predictions: indeed we expect that $\delta \propto 1/n$.
In Fig.~\ref{fig:deltan} 
we have plotted the function $\delta(n)$ given in Table~1, together
with a best fit to a function proportional to $1/n$. 
For the lattice parameters used in the simulation 
we see that $\delta(n)$ behaves as
expected for small values of $n$, while the relation $\delta \propto 1/n$
breaks down for $n \gtrsim 30-40$, indicating that
the picture of the standard cooling as a diffusion
process may work well only for moderately small values of $n$.

\section{The topological susceptibility}
\label{sec:topsusc}

In this Section we apply the new cooling method to the determination
of the physical topological susceptibility and compare the results to those from
an alternative approach, the field theoretical
method~\cite{CDPV90} combined with smearing~\cite{CDPV96}. As for the comparison
of the new cooling with the standard and Pisa coolings, we have already shown
in the previous Section that they give equivalent results for $\chi^L$.

The lattice discretization of the topological charge density and of the 
topological susceptibility has already been discussed in
Section~\ref{sec:def}. 
In the relation existing between the lattice topological susceptibility
$\chi^L$ and the continuum one, Eq.~(\ref{reneq}),
$M(\beta)$ indicates the mixings of $\chi^L$ to operators with
the same quantum numbers. More precisely one can write
\begin{equation}
\label{reneq3}
M(\beta) = A(\beta)\,a^2 \langle T\rangle_{\rm np} + P(\beta)\langle I \rangle 
+ O(a^4)\;.
\end{equation} 
The first and second terms in the r.h.s. are the mixings with 
the trace of the energy--momentum tensor 
and with the unit operator respectively (``np'' means the purely
non--perturbative part of $\langle T\rangle$).
 
The mixing coefficients $A(\beta)$ and $P(\beta)$ as well as the
multiplicative renormalization $Z(\beta)$ can be calculated 
in perturbation theory. The perturbative series for $A(\beta)$
starts from the order $1/\beta^3$ and arguments can be 
given~\cite{FP94} which justify that 
$A(\beta)\,a^2 \langle T\rangle_{\rm np}$
is safely negligible in the scaling window of the simulation. 
In Ref.~\cite{FP93} the 
perturbative series for $Z(\beta)$ has been 
calculated up to the order $1/\beta^2$ and
the perturbative series of $P(\beta)$ at the first two non--zero
orders $1/\beta^4$ and $1/\beta^5$. On the other hand there are no available
perturbative estimates of the $O(a^4)$ terms in Eq.(\ref{reneq3}).

However, a more powerful, purely numerical technique can be used to  
get a non--perturbative determination of $Z(\beta)$ and of the whole additive
renormalization $M(\beta)$. This technique is 
the so--called ``heating method''~\cite{DV92}.

The idea of the heating method is to determine the average values
of topological quantities on samples of configurations
obtained by thermalizing the short--range fluctuations, which are responsible
for the renormalizations, on configurations of well defined
topological background. For instance if one applies 
several thermalization steps at a given value of $\beta$
on a configuration containing one
discretized instanton of charge $Q = 1$ and measures the average
value of $Q^L$, then the renormalization constant $Z(\beta)$ can be 
determined as $Z(\beta)=\langle Q^L\rangle/Q$ (here 
$\langle\cdot\rangle$ indicates an average in the background of a
fixed charge $Q$). Analogously, by thermalizing
a trivial configuration (for example a configuration where for all lattice sites 
$x$, the respective fields are $z(x)=(1,0,0,\dots)$ and
$\lambda_\mu(x)=1$) at a given $\beta$ and measuring
the average value of $\chi^L$, one gets a determination of $M(\beta)$.
In the first case one obtains a non--perturbative estimate of 
$Z(\beta)$ as long as the $\overline{\hbox{MS}}$ value of the
background topological charge $Q$ is chosen
(see Section~\ref{sec:def}). In the second case the
method amounts to impose the requirement that the physical
susceptibility $\chi$ vanishes in the absence of instantons.

The heating method has been extensively applied in the following.
Once all the renormalization constants are known, the extraction of
$\chi$ from Eq.~(\ref{reneq}) is possible and the result is
called the field theoretical determination.

\begin{figure}[tb]
\begin{center}
\hspace{0.3cm}
\includegraphics[width=0.65\textwidth]{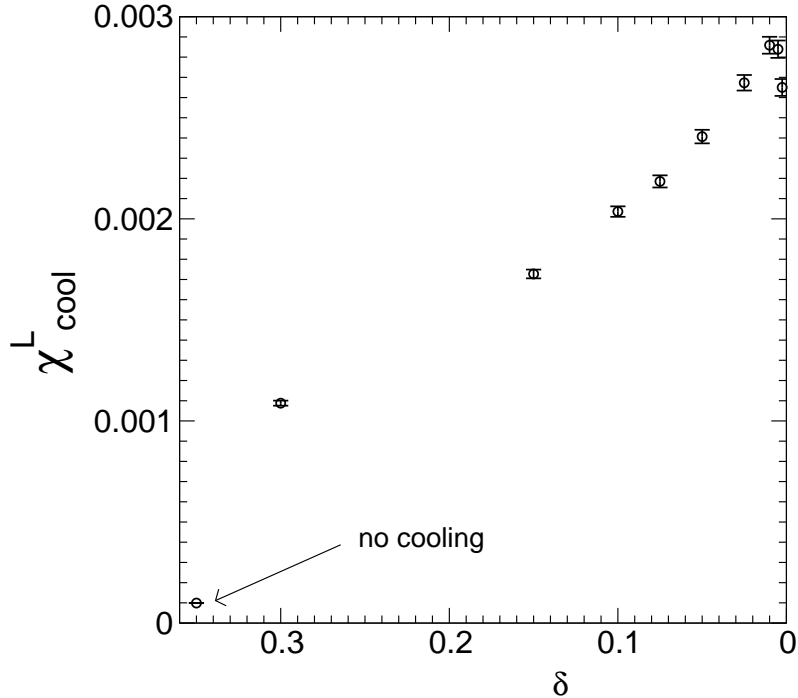}
\end{center}
\caption{$\chi^L_{\mathrm{cool}}$ as a function of the $\delta$
parameter of the new cooling for CP$^9$ at $\beta=0.70$ on a $24^2$ lattice.}
\label{fig:cool,smear}
\end{figure}

Since the continuum $\chi$ is extracted from the lattice $\chi^L$
by subtracting the renormalization effects, 
it happens that, if $M(\beta)$
is a large part of the whole lattice signal and $Z(\beta)$ is small,
the physical signal for $\chi$ is extracted with very large errors. This
is the case when one uses the lattice discretization of Eq.~(\ref{q^L}).
However, since both $Z(\beta)$ and $M(\beta)$ depend on the  
discretization $q^L(x)$, one can exploit the arbitrariness in the
lattice definition and use an improved operator.
To this aim, following the idea of Ref.~\cite{CDPV96}, already used for the
determination of $\chi$ in Yang--Mills theories~\cite{ADD97}, we have used a smeared 
topological charge density 
operator, which is built from the original
operator $q^L(x)$, defined in Eq.~(\ref{q^L}), by replacing the fields 
$z(x)$ and $\lambda_\mu(x)$ with
\begin{displaymath}
\label{smear}
z^{\mathrm{smear}}(x) = {\mathcal{N}}_z \biggl[ (1-c) z(x) + \frac{c}{4} 
\sum_{\mu=1,2} \biggl( z(x-\widehat\mu) \, 
 \lambda_\mu(x-\widehat\mu)
+ z(x+\widehat\mu) \, \bar \lambda_\mu(x) \biggr)
\biggr] \;, \;\;\;\;
\end{displaymath}
\begin{equation}
\lambda_\mu^{\mathrm{smear}}(x) = 
{\mathcal{N}_\lambda} \biggl[ (1-c) \lambda_\mu(x) + c \, 
\frac{\bar z(x) \cdot z(x+\widehat\mu)}{||\bar z(x) \cdot z(x+\widehat\mu)||}
\biggr] \,,
\label{smeardef}
\end{equation}
where ${\mathcal{N}}_z$, ${\mathcal{N}_\lambda}$ are normalization constants
which allow that $||z^{\rm smear}(x)|| = ||\lambda_\mu^{\rm smear}(x)||=1$ 
and $c$ has been chosen to be equal to 0.65$\;$\footnote{This
choice has been based on the fact that {\it i)} it is convenient to
take $c$ as large as possible and {\it ii)} 
for $c > 2/3$ the smearing
procedure behaves in a radically different way, leading, when iterated,
to a completely disordered field configuration instead of a smoother
one. A similar behavior has been observed in Yang--Mills theories~\cite{hetrick}.}. 
The smearing can be iterated 
at will by defining the $n$--th level of smeared fields from the
$(n-1)$--th level in an analogous fashion as shown in Eq.~(\ref{smeardef}). For 
each level of smearing a relation like Eq.~(\ref{reneq}) holds although
$Z(\beta)$ and $M(\beta)$ change.
Using smeared operators these renormalization constants
get closer to 1 and 0, respectively, as the smearing 
level increases, thus allowing a much more precise determination of $\chi$.

We performed a numerical simulation for the CP$^9$ model at several
values of $\beta$ and $L$. We used the same 
updating algorithm of the previous Section.
For most simulations we collected 100K
equilibrium configurations after discarding 10K configurations to
allow thermalization. The successive 
equilibrium configurations were decorrelated by 10 updating steps.
By averaging on the thermal equilibrium ensemble we determined the unsmeared 
lattice topological susceptibility $\chi^L$ and the smeared ones, 
$\chi^L_{\mathrm{smear}}$, for 1 to 10 smearing levels.
These results, together with $Z(\beta)$ and $M(\beta)$ coming from the 
heating method, allow to extract $a^2 \chi$ 
from Eq.~(\ref{reneq}). 

Every 50 updating steps we have also applied the new cooling 
method at several values of the $\delta$ parameter 
ranging from 0.3 to $5\times 10^{-4}$. Since cooling eliminates the 
short--distance fluctuations, the value of $\chi^L$ after cooling,
hereafter called $\chi^L_{\mathrm{cool}}$, can be
written as in Eq.~(\ref{reneq}) with $Z\approx 1$ and 
$M\approx 0$ if the cooling has been protracted enough.
Hence it should provide directly $a^2 \chi$. 
Consistency requires that the cooling and the field theoretical
methods provide the same result.

\begin{figure}[tb]
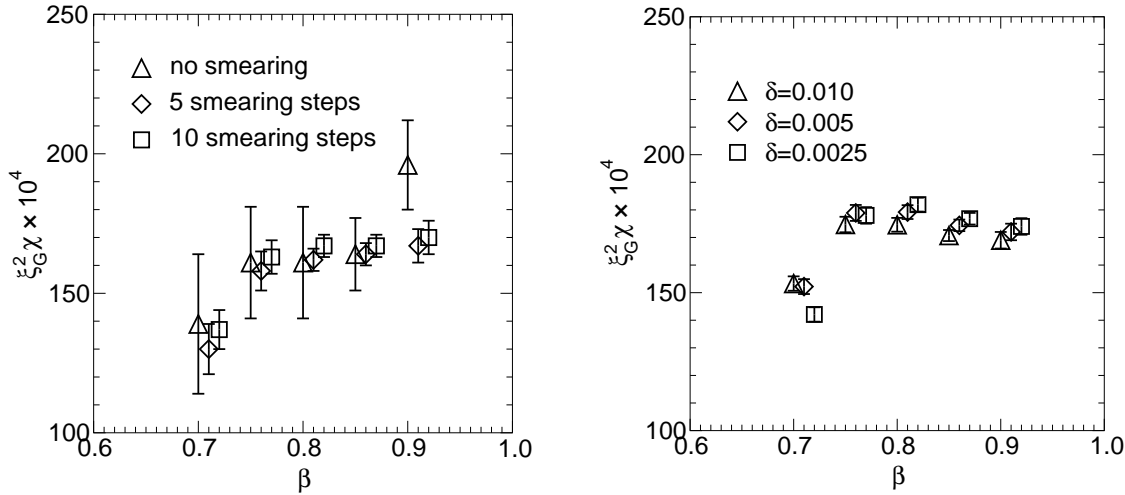

\begin{minipage}{0.48\textwidth}
\begin{center}
\includegraphics[width=0.9\textwidth]{scaling_ft_cp9.eps}
\end{center}
\end{minipage}
\begin{minipage}{0.48\textwidth}
\begin{center}
\includegraphics[width=0.9\textwidth]{scaling_cool_cp9.eps}
\end{center}
\end{minipage}
\caption{Scaling of the topological susceptibility of CP$^9$ from the field 
theoretical with smearing (left) and from the new cooling (right). Data points
have been slightly shifted along the $\beta$ axis for the sake of 
readability.}\label{fig:scal}
\end{figure}

The summary of the performed simulations is presented in Table~2. 
In Fig.~\ref{fig:cool,smear} we show the results for $\chi^L_{\rm cool}$
from the new cooling at several values for the $\delta$
parameter at $\beta=0.70$ on a $24^2$ lattice. 
We interpret the behavior under the new cooling in the following way: as
$\delta$ decreases, the cooling erases an even larger amount of
quantum fluctuations thus bringing $\chi^L_{\rm cool}$ to a maximum. Afterwards,
for too small values of $\delta$, the cooling begins to affect also the 
topological fluctuations and part of the topological signal gets
lost. These considerations 
suggest that a convenient choice of the $\delta$ parameter may be around
the value for which $\chi^L_{\rm cool}$ reaches the maximum. At this
value of $\delta$ we then assume that $Z\approx 1$ and $M\approx 0$.
The choice of a smaller value for $\delta$
would not be dramatically dangerous, but would push the scaling
region towards larger values of $\beta$.

In Table~3 we summarize the results for $\chi^L$ and for 
$\xi_G^2\,\chi$ obtained for CP$^9$ by the 
field theoretical method combined with smearing 
(0, 5 and 10 levels). In Table~4 the results for
$\chi^L_{\mathrm{cool}}$ and $\xi^2_G\,\chi$
obtained by the new cooling method for $\delta$ around the 
peak value (see Fig.~\ref{fig:cool,smear} for $\beta=0.70$) are
shown. Fig.~\ref{fig:scal} displays the scaling
of $\xi_G^2\,\chi$ for the cases of 0, 5 and 10 smearing levels 
and the field theoretical method (left) and with the new cooling method
for $\delta$=0.0100, 0.0050, 0.0025 (right). From this figure we see
that there is practically no dependence
on the smearing level, 
although error bars strongly decrease with the number of smearing 
levels. As for the cooling method, we see that there is a satisfying consistence 
among the results for the different values of $\delta$ and also between these 
results and those from the field theoretical method.

\section{Summary and conclusions}
\label{sec:summary}

The study of the topological properties of the vacuum of a
field theory simulated on a lattice is made difficult by the presence
of quantum fluctuations which hinder the extraction of the relevant physical
information. Among other methods, cooling has been proposed as a
technique to wash out such fluctuations and reveal the topological
background of any single field configuration. In the present paper 
we have made a comparison among three different cooling methods 
using as test--field the 2d CP$^{N-1}$ model defined in 
Section~\ref{sec:def}.

The three cooling methods under study have been defined in 
Section~\ref{sec:cooling}. They are: the standard cooling (a local
minimization of the Euclidean lattice action), the Pisa cooling (the
same but with local minimizations constrained to be smaller than a
given bound) and a new cooling (recently introduced in
Ref.~\cite{PPS99}) where the local minimizations are accepted only if they
are larger than a given bound.

Firstly, we have studied the performance of the three coolings on
classical non--equi-\break librium configurations representing  
1--instanton solutions and instanton--anti-instanton
pairs. We have placed one such object on the lattice
and have performed a series of cooling iterations in order to measure 
several physical (topological and non--topological) observables.
In all situations the three coolings have yielded the same result 
---see Figs.~\ref{fig:Q,S_vs_rho} and~\ref{fig:Qcorr_vs_S}.

After these results, we have assumed that the three cooling methods
are equivalent and that a correspondence can be established between
them in the sense that a given number of iterations of the Pisa or
the standard coolings corresponds to a precise value for the bound in 
the new cooling if the value of the energy $E$, Eq.~(\ref{energy}),
is the same after applying the coolings. This correspondence has been used
in the subsequent investigation and, for the CP$^3$ model at
$\beta=1.05$ on a $76^2$ lattice, it is shown in Table~1.

We have studied the performance of the cooling methods on a set
of equilibrium configurations obtained after a Monte Carlo simulation. 
On these thermalized configurations we have first extracted several
physical quantities after applying cooling: the lattice topological
susceptibility is the same when equivalent amounts of cooling, in the sense
of Table~1, are applied ---see
Fig.~\ref{fig:chiconfr}. The same happens for the magnetic susceptibility
and for the correlation length as far as the cooling iteration is not
protracted too much ---see Fig.~\ref{fig:susc,csi}.
The same agreement is seen for the shell correlation function 
$\langle G(r)\rangle$ if $r/a$ is large enough ---see Fig.~\ref{fig:q_corr}.
Then we have peered the action density and the topological charge 
density distributions obtained with the application of equivalent amounts
of cooling, in the sense of Table~1. In 
all cases we have again obtained completely analogous distributions
---see Fig.~\ref{fig:therm} for an example.
These results altogether strongly suggest that the three cooling methods,
although different in the procedure, behave equivalently.

By using the correspondence given in Table~1 we have tested
the picture of cooling as a diffusion process. For the lattice
parameters used in this analysis, this picture works
well only for a moderately small number of iterations ($n \lesssim 40$).

Finally we have compared the results obtained for the topological susceptibility
in the CP$^9$ model by the new cooling method with those 
extracted from a well--tested method: the so--called field theoretical 
method improved with smearing.
In Tables~3 and~4 we give the results for the quantity
$\xi_G^2\,\chi$. They agree fairly well among them, see
Fig.~\ref{fig:scal}, and also with the large $N$
estimate~\cite{camporossi} which provides $\xi_G^2\,\chi=153$
up to order $O(1/N^2)$. 

We used the standard action both 
to generate thermal equilibrium configurations
in Monte Carlo simulations and during the cooling procedure. 
We expect that the above results concerning the equivalence between
the different cooling techniques should not depend strongly on
the action used during cooling.
However the comparison of the three cooling
techniques by using different lattice actions
is worth to be pursued in a future work.

\section*{Acknowledgement}
We would like to thank P.~Cea, Ph.~de~Forcrand, A.~Di~Giacomo, P.~Rossi,
I.O.~Stamatescu and E.~Vicari for useful discussions.

\newpage

\centerline{\bf TABLES}

\begin{table}[hbtp]
\setlength{\tabcolsep}{1.00pc}
\centering
\caption{Relation between the new cooling parameter $\delta$ and 
the number of standard cooling iterations $n$, at $\beta = 1.05$ 
on a $76^2$ lattice in the CP$^{3}$ model.
The errors given in the second column derive from the uncertainty
in the matching procedure.}
\begin{tabular}{cc}
\hline
\hline
$n$ (standard cooling) & $\delta$ (new cooling) \\ 
\hline
3 & 0.068240(50)   \\
4 & 0.048500(50)   \\
5 & 0.038000(40)   \\
7 & 0.026800(30)   \\
10 & 0.018840(40)  \\
15 & 0.012850(40)  \\
20 & 0.009925(40)  \\
30 & 0.007000(30)  \\
40 & 0.005492(20)  \\
50 & 0.004602(20)  \\
60 & 0.003990(30)  \\
70 & 0.003530(30)  \\
100 & 0.002720(20) \\
150 & 0.002038(10) \\
200 & 0.001662(10) \\
\hline
\hline
\end{tabular}
\end{table}

\begin{table}[hbtp]
\setlength{\tabcolsep}{1.88pc}
\centering
\caption{Summary of the simulations for the CP$^3$ and CP$^9$ models.}
\begin{tabular}{ccccll}
\hline
\hline
$N$ & $\beta$ & $L$ & stat & $\ \ \ \ \ E$ & $\ \ \ \xi_G/a$ \\ 
\hline
 4 & 1.05 &  76 & 100k & 0.530638(12)  &  7.688(29)    \\
\hline
10 & 0.70 &  24 & 100k & 0.78429(3)   &  2.316(3)      \\
10 & 0.75 &  30 & 10k & 0.72013(8)   &  3.297(11)     \\
10 & 0.75 &  32 & 100k & 0.720146(25) &  3.2848(24)    \\
10 & 0.75 &  60 & 10k & 0.72024(5)   &  3.248(25)     \\
10 & 0.80 &  48 & 100k & 0.667000(15) &  4.608(4)      \\
10 & 0.85 &  64 & 100k & 0.622271(10) &  6.419(7)      \\
10 & 0.85 &  80 & 10k & 0.622357(29) &  6.371(29)     \\
10 & 0.90 &  90 & 100k & 0.583830(7)  &  8.836(10)     \\
\hline
\hline
\end{tabular}
\end{table}

\begin{table}[hbtp]
\setlength{\tabcolsep}{0.69pc}
\centering
\caption{Summary of the results for the topological susceptibility
of the CP$^9$ model by the field theoretical method at various smearing levels.}
\begin{tabular}{cccccccc}
\hline
\hline
$\beta$ & $L$ & level & $\chi^L\times 10^4$ & $M(\beta)\times 10^6$ & $Z(\beta)$ &
$a^2\, \chi\times 10^4$ & $\xi_G^2\,\chi\times 10^4$ \\
\hline
     &    &  0 &  0.992(5)  & 33(3)   & 0.160(10) & 26(5)     & 139(25) \\
0.70 & 24 &  5 &  16.72(8)  & 78(15)  & 0.810(20) & 24.3(1.6) & 130(9)  \\
     &    & 10 &  21.40(11) & 45(15)  & 0.907(15) & 25.5(1.2) & 137(7)  \\
\hline
     &    &  0 &  0.813(12) & 23.0(1.0) & 0.197(10) & 15.0(2.1) & 163(24) \\
0.75 & 30 &  5 &  10.41(16) & 16.0(2.5) & 0.837(15) & 14.6(8)   & 159(10) \\
     &    & 10 &  12.60(19) & 5.5(1.0)  & 0.915(15) & 15.0(7)   & 163(9)  \\
\hline
     &    &  0 &  0.809(4)  & 23.0(1.0) & 0.197(10) & 15.0(2.1) & 161(20) \\
0.75 & 32 &  5 &  10.43(5)  & 16.0(2.5) & 0.837(15) & 14.6(8)   & 158(7)  \\
     &    & 10 &  12.67(6)  & 5.5(1.0)  & 0.915(15) & 15.0(7)   & 163(6)  \\
\hline
     &    &  0 &  0.810(12) & 23.0(1.0) & 0.197(10) & 15.0(2.1) & 158(24) \\
0.75 & 60 &  5 &  10.51(15) & 16.0(2.5) & 0.837(15) & 14.8(8)   & 156(11) \\
     &    & 10 &  12.82(19) & 5.5(1.0)  & 0.915(15) & 15.2(7)   & 161(10) \\
\hline
     &    &  0 &  0.5981(20) & 16.0(1.5) & 0.240(10) & 7.6(9)   & 161(20) \\
0.80 & 48 &  5 &  5.87(3)    & 2.0(6)    & 0.876(8)  & 7.63(19) & 162(4)  \\
     &    & 10 &  6.87(4)    & 0.40(15)  & 0.933(7)  & 7.88(17) & 167(4)  \\
\hline
     &    &  0 &  0.4208(28) & 13.0(7)   & 0.270(6)  & 4.0(3)   & 164(13) \\
0.85 & 64 &  5 &  3.180(27)  & 0.40(20)  & 0.892(7)  & 3.99(10) & 164(4)  \\
     &    & 10 &  3.61(3)    & 0.05(3)   & 0.945(7)  & 4.05(9)  & 167(4)  \\
\hline
     &    &  0 &  0.423(8)   & 13.0(7)   & 0.270(6)  & 4.0(4)   & 163(17) \\
0.85 & 80 &  5 &  3.19(8)    & 0.40(20)  & 0.892(7)  & 4.00(17) & 162(8)  \\
     &    & 10 &  3.63(9)    & 0.05(3)   & 0.945(7)  & 4.06(17) & 165(8)  \\
\hline
     &    &  0 &  0.2975(15) & 9.5(5)    & 0.284(7)  & 2.51(20) & 196(16) \\
0.90 & 90 &  5 &  1.74(3)    & 0.10(3)   & 0.902(8)  & 2.14(7)  & 167(6)  \\
     &    & 10 &  1.94(3)    & 0.010(3)  & 0.945(9)  & 2.18(8)) & 170(6)  \\
\hline
\hline
\end{tabular}
\end{table}

\begin{table}[hbtp]
\setlength{\tabcolsep}{2.27pc}
\centering
\caption{Summary of the results for the topological susceptibility of the 
CP$^9$ model from the new cooling method at various $\delta$ values.}
\begin{tabular}{ccccc}
\hline
\hline
$\beta$ & $L$ & $\delta$ & $\chi^L_{\mathrm{cool}}\times 10^4$ & 
$\xi_G^2\,\chi\times 10^4$ \\
\hline
     &    & 0.0500 &  24.1(3) & 129.0(2.1) \\
     &    & 0.0250 &  26.7(4) & 143.3(2.4) \\
0.70 & 24 & 0.0100 &  28.6(4) & 153.3(2.6) \\
     &    & 0.0050 &  28.4(4) & 152.2(2.7) \\
     &    & 0.0025 &  26.5(4) & 142.1(2.6) \\
\hline
     &    & 0.0500 & 13.5(5) & 146(6) \\
     &    & 0.0250 & 14.8(5) & 161(6) \\
0.75 & 30 & 0.0100 & 15.9(5) & 173(7) \\
     &    & 0.0050 & 16.2(5) & 177(7) \\
     &    & 0.0025 & 16.1(5) & 175(7) \\
\hline
     &    & 0.0500 &  13.70(20) & 147.8(2.4) \\
     &    & 0.0250 &  15.08(22) & 162.7(2.6) \\
0.75 & 32 & 0.0100 &  16.19(24) & 174.7(2.8) \\
     &    & 0.0050 &  16.58(24) & 178.9(2.9) \\
     &    & 0.0025 &  16.51(25) & 178(3)     \\
\hline
     &    & 0.0500 &  13.5(4) & 143(7) \\
     &    & 0.0250 &  14.8(5) & 156(7) \\
0.75 & 60 & 0.0100 &  15.9(5) & 167(8) \\
     &    & 0.0050 &  16.4(5) & 173(8) \\
     &    & 0.0025 &  16.6(5) & 175(8) \\
\hline
     &    & 0.0500 &  7.12(9)  & 151.1(2.1)   \\
     &    & 0.0250 &  7.73(9)  & 164.1(2.3)   \\
0.80 & 48 & 0.0100 &  8.22(10) & 174.6(2.5)   \\
     &    & 0.0050 &  8.44(10) & 179.2(2.5)   \\
     &    & 0.0025 &  8.57(10) & 181.9(2.5)   \\
\hline
     &    & 0.0500 & 3.68(4) & 151.8(1.9) \\
     &    & 0.0250 & 3.94(4) & 162.5(2.0) \\
0.85 & 64 & 0.0100 & 4.14(4) & 170.7(2.0) \\
     &    & 0.0050 & 4.23(4) & 174.4(2.1) \\
     &    & 0.0025 & 4.29(4) & 176.8(2.1) \\
\hline
     &    & 0.0500 & 3.63(12) & 147(6) \\
     &    & 0.0250 & 3.89(13) & 158(7) \\
0.85 & 80 & 0.0100 & 4.09(14) & 166(7) \\
     &    & 0.0050 & 4.17(14) & 169(7) \\
     &    & 0.0025 & 4.23(14) & 172(7) \\
\hline
     &    & 0.0500 & 1.97(3) & 153(3) \\
     &    & 0.0250 & 2.08(4) & 162(3) \\
0.90 & 90 & 0.0100 & 2.17(4) & 169(3) \\
     &    & 0.0050 & 2.21(4) & 172(3) \\
     &    & 0.0025 & 2.23(4) & 174(3) \\
\hline
\hline
\end{tabular}
\end{table}

\end{document}